\documentclass[11pt]{article} 
\usepackage{latexsym} %
\input{epsf.sty}
\newcommand{\UGR}{\em Dpto. de F\'{\i}sica Te\'orica y del Cosmos,  
		       Universidad de Granada, 18071 Granada, Spain}   
\newcommand{\UP}{\em Dip. di Fisica, 
		       Universit\'a di Padova, 35131 Padova, Italy}

\newcommand{\be}{\begin{equation}}
\newcommand{\ee}{\end{equation}}
\newcommand{\bea}{\begin{eqnarray}}
\newcommand{\eea}{\end{eqnarray}}
\newcommand{\ba}{\begin{array}}
\newcommand{\ea}{\end{array}}
\newcommand{\bfi}[1]{\begin{figure}[#1]}
\newcommand{\efi}{\end{figure}}
\newcommand{\bpi}[2]{\begin{picture}(#1,#2)}
\newcommand{\epi}{\end{picture}}

\newcommand{\g}{\gamma}
\newcommand{\prop}{\Delta}
\newcommand{\propm}{{\Delta}_m}
\newcommand{\dsl}{\not \! \partial}
\renewcommand{\d}{\partial}
\newcommand{\T}{\mathrm{T}}
\newcommand{\B}{\mathrm{B}}

\newcommand{\I}{\mathrm{I}}
\newcommand{\J}{\mathrm{J}}

\def\QQa{\renewcommand{\baselinestretch}{1.3}\Huge\large\normalsize}

\makeatletter 
\def\secteqno{\@addtoreset{equation}{section}%
\def\theequation{\thesection.\arabic{equation}}} 
 
\def\endsecteqno{\def{theequation\{\@ifundefined{chapter}%
{\arabic{equation}}{\thechapter.\arabic{equation}}}} 
\makeatother

\begin{document}

\pagestyle{empty} 
 \begin{flushright}  UG-FT-71/96 \\
	             DFPD 97/TH/07 \\
		     February 1997
  \end{flushright} 
	      
\vspace*{2cm}                               
\begin{center} 
\large{\bf Supergravity corrections to $(g-2)_l$ in 
differential renormalization } 
\vskip .6truein 
\centerline {F. del \'Aguila, $^{a,}$\footnote{e-mail: faguila@goliat.ugr.es}
             A. Culatti, $^{a,b,}$\footnote{e-mail:  culatti@mvxpd5.pd.infn.it}
             R. Mu\~noz Tapia, $^{a,}$\footnote{e-mail: rmt@ugr.es} and
             M. P\'erez-Victoria $^{a,}$\footnote{e-mail: mpv@ugr.es}}  
\end{center} 
\vspace{.3cm} 
\leftline 
{$^{a}$\UGR} 
\vspace{.1cm} 
\leftline
{$^{b}$\UP} 
\vspace{1.5cm} 
  
\centerline{\bf Abstract} 
\medskip 

The method of differential renormalization is extended to the calculation of the
one--loop 
gra\-vi\-ton and gravitino corrections to $(g-2)_l$ in unbroken supergravity. 
Rewriting the singular contributions of all the diagrams in terms of only one 
singular function, $U(1)$ gauge invariance and supersymmetry are preserved. 
We compare 
this calculation with previous ones which made use of momentum space 
regularization (renormalization) methods.

\newpage
\pagestyle{plain}
\QQa
\secteqno


\section{Introduction}

The method of differential renormalization (DR) has appeared 
recently~\cite{FJL}.
It is a method of renormalization in coordinate space that
yields directly finite amplitudes and leaves the space--time
dimension unchanged. Thus it is expected to be especially 
suited for supersymmetric theories, where the dimension of the 
space-time appears to be crucial for preserving the symmetry.
DR has already been 
successfully applied in several contexts: the Wess--Zumino 
model~\cite{SusyDR}, 
lower--dimensional~\cite{lowDR} and non--abelian~\cite{nonabDR} 
gauge theories,
two--loop QED~\cite{QEDDR}, chiral models~\cite{chiralDR}, 
non--relativistic anyon models~\cite{anyonDR} and curved 
space--time~\cite{curvedDR}. 
Other formal aspects of the method 
have been developed in~\cite{SistDR,MassiveDR} and  
different versions of DR can be found in~\cite{OtroDR}. 
It is 
the purpose here to push even further the method and tackle 
a relatively complex problem such as the computation of $(g-2)_l$ 
in supergravity (SUGRA). 
This is the first time  that DR is applied to the calculation
of a physical observable and to the calculation of gravitational 
corrections. As we will show our
results  preserve supersymmetry (SUSY) and abelian gauge invariance. 

Although SUSY is not an
exact symmetry of Nature, it is believed
that any fundamental theory must be originally supersymmetric.
It is also known that when SUSY is made local it naturally
includes gravity~\cite{sugra1}. The resulting theory is 
non--renormalizable 
but it is constrained by the symmetries. 
One of the few finite calculations in gravity is the one--loop 
correction to the anomalous magnetic moment of the
lepton $(g-2)_l$ \cite{figrco}. 
In SUGRA not only is $(g-2)_l$ finite but SUSY requires it to 
be zero \cite{amm}. 
Therefore, the anomalous magnetic moment of the
lepton besides being an observable, is also an ideal arena 
to check theoretical implications and to perform consistency tests 
of the methods of regularization in SUGRA. 

In a supersymmetric theory $(g-2)_l$ 
vanishes because no such term appears in the Lagrangian of a chiral 
supermultiplet~\cite{amm}. (This has been generalized to a set of 
sum rules  valid for any charged supersymmetric 
multiplet~\cite{sumrules,susyver}.) 
Hence, as long as SUSY is preserved, all quantum
corrections must cancel order by order.	
Ferrara and Remiddi also proved explicitly  that in global SUSY the 
one--loop QED corrections, order $e^3$, do cancel. In this case two diagrams
contribute (see Fig. 1), one exchanging a photon and the lepton,
and the other a photino and the corresponding sleptons, and both are finite. 


\begin{figure}[thb]
\setlength{\unitlength}{1cm}
\begin{center}
\begin{picture}(8,4)
\epsfxsize=10cm
\put(0,0){\epsfig{file=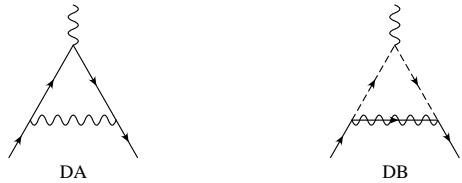,width=6cm}}
\end{picture}
\end{center}
\caption{Diagrams of order $e^3$ contributing to $(g-2)_l$ 
in superQED. \label{figqed}}
\end{figure}

\begin{figure}[htb]
\setlength{\unitlength}{1cm}
\begin{center}
\begin{picture}(15,6)
\put(0,0){\epsfig{file=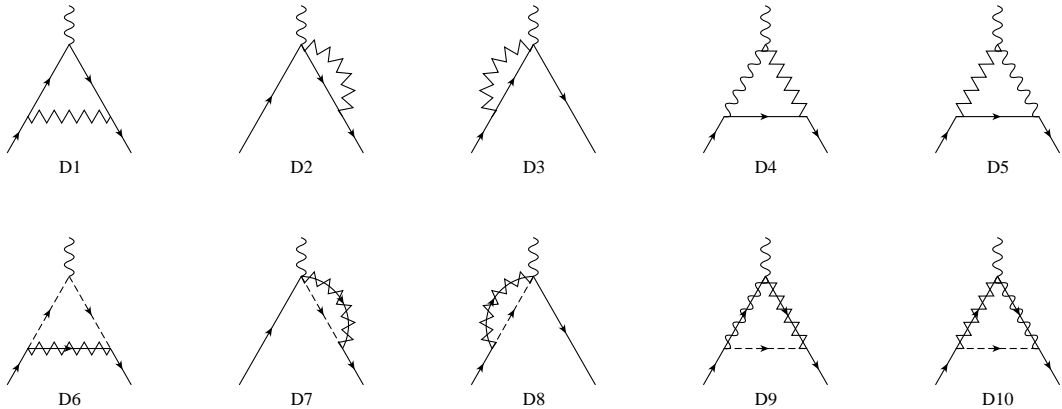,width=14cm}}
\end{picture}
\end{center}
\caption{Diagrams of order $e \kappa^2 $ contributing to $(g-2)_l$ 
in SUGRA. A graviton is exchanged in diagrams D1-D5 and a gravitino
in D6-D10. \label{figsugra}}
\end{figure}

The one--loop gravitational
corrections are of order $e\kappa^2= 8\pi e G_N$, resulting from
a graviton or gravitino exchange. Using dimensional
regularization~\cite{dimreg},
Berends and Gastmans calculated  the five diagrams where a graviton
is exchanged (see Fig. 2)\cite{figrco}. All five diagrams are infinite but
their sum is finite. The finiteness of $(g-2)_l$ 
in a non--renormalizable theory such as gravity
seemed miraculous. Del Aguila et al.~\cite{Paco} and 
Bellucci et al.~\cite{Bellucci}
checked that when gravitation is embedded in a supersym\-metric theory
(unbroken), the
contributions from the graviton and the gravitino cancel, as 
required by SUSY. 
Bellucci et al. also 
traced back to an effective chiral symmetry in the 
gravitino sector the finiteness of the gravitino contribution 
and then of the graviton contribution, if their sum has to vanish. 
Dimensional regularization does not yield a vanishing 
value for $(g-2)_l$. This is not
so surprising for dimensional regularization is known to break
SUSY. A (one--loop) SUSY preserving method such as
dimensional reduction~\cite{dimred} is then required in order to obtain 
such a cancellation and this 
was shown to be the case. However, whether the finite
contributions to $(g-2)_l$ of the graviton and gravitino sectors 
are well--defined quantities for some unknown reason (symmetry) is a 
question which remained unanswered. 
The lack of manageable, alternative regularization 
methods preserving SUSY has prevented to address this question although 
the answer is expected to be negative as hinted
in~\cite{Bellucci,Wilcox,figrco}. 

In this paper we use DR to calculate the one--loop
corrections to $(g-2)_l$ in unbroken SUGRA.  
We find that in DR the individual
contributions of the graviton and gravitino 
sectors, although finite and opposite, are different from previous 
results. 
Thus, although SUSY is preserved, these two contributions are
not well--defined separately, only their sum is.

The method of DR is based on the observation that in coordinate space
singularities arise when points in the quantum fields coincide. 
The idea is to write singular expressions as derivatives of 
less singular functions. One  solves some
differential equations and considers the solutions as a definition
of the amplitudes in the sense of distributions, {\it i.e.}, the 
derivatives are  understood to act on test functions. 
In fact the method gives a definition of the expressions
at the ill--defined  points.
The constants that naturally appear in the differential equations
play the role of renormalization scales.
The resulting finite amplitudes satisfy the renormalization
group equations.
While retaining the spirit of the method, we have developed
some aspects that may be considered unsatisfactory in the original
version. For instance, 
gauge invariance had to be checked at each step of the computation
and the renormalization scales chosen accordingly. In this 
paper we show that it is possible to preserve gauge invariance
(no term proportional to the photon momentum is generated at one loop)
and supersymmetry 
($(g-2)_l$ has no one--loop corrections) 
if related singularities are treated always in the 
same way throughout
the computations. Only one scale for each type of singularity 
is then needed. In particular, for the computation of $(g-2)_l$
only the scale corresponding to logarithmic singularities appears.
Obviously it cancels out  when
all diagrams are summed up.
These new features  
of the DR approach  will be more 
extensively discussed
in a simpler context in a forthcoming publication \cite{SQED}.

The plan of the paper is as follows.
In Section \ref{SUGRAlagrangian} the lagrangian describing a minimal 
superQED--SUGRA theory is presented.
Section \ref{procedure}
is devoted to describing the constrained differential 
renormalization procedure which we will follow.
Section \ref{example} contains 
the detailed calculation of one diagram, 
where the main techniques 
used in the paper are applied, whereas in Section  \ref{complete} 
we discuss briefly the calculation of the remaining 
diagrams contributing to $(g-2)_l$ and give the final results. 
We finish with the conclusions in Section \ref{conclusions}. 
Three Appendices gather 
the Feynman rules in Euclidean space and other technical details. 


\section{The SUGRA Lagrangian} \label{SUGRAlagrangian}
The coupling of matter to supergravity has been extensively studied in the
literature~\cite{sugra2}. We are interested here only in first--order 
gravitational
corrections. We shall use the  lagrangian of superQED--SUGRA 
obtained by
imposing canonical kinetic terms (minimal K\"ahler potential and $f$ function) 
and expanding the curved metric around the Minkowski one ($\eta_{\alpha\beta}$).
The interaction lagrangian in Minkowski space reads 
\bea
   {\cal L}_{ee\g+\tilde{e}\tilde{e}\g+e\tilde{e}\tilde{\g}} & = &  
    - e \bar{\Psi} \not \! \! A \Psi - [ i e A^\mu \phi_{L}^\dagger 
    \stackrel{\leftrightarrow}{\d}_\mu \phi_{L}  \nonumber \\
    & &  \mbox{} - e \sqrt{2}(\bar{\lambda} \phi_L^\dagger
    {\cal P}_L \Psi  +
    \mathrm{h.c.}) +(L \leftrightarrow R) ] ~, 
    \label{lag1} \\
   {\cal L}_{eeg+eeg\g} & = & 
    -\frac{ \kappa}{4}  h^{\alpha\beta} [(i \bar{\Psi} (\g_\alpha \d_\beta
    + \g_\beta \d_\alpha) \Psi + \mathrm{h.c.})  \nonumber \\
    & &  \mbox{} - 2 e \bar{\Psi}(\g_\alpha A_\beta + 
    \g_\beta A_\alpha) \Psi ] ~,\\
   {\cal L}_{e \tilde{e} \tilde{g} + e \tilde{e} \tilde{g} \g} & = &
     - \frac{\kappa}{\sqrt{2}} [ \bar{\chi}^\nu 
     {\cal P}_{L} (i \dsl -m) \phi_{L}^\dagger
     \g_\nu \Psi   \nonumber \\
     & &  \mbox{} + e \bar{\chi}^\nu  {\cal P}_{L} 
     \not \! \! A  \phi_{L}^\dagger  \g_\nu  \Psi
     + \mathrm{h.c.} ] + (L \leftrightarrow R) ~, \\
   {\cal L}_{\g \g g + \g \tilde{\g} \tilde{g}} & = &
     \kappa [ h^{\alpha \beta} (F_{\alpha\mu} F_\beta^{~\mu} 
     - {1 \over 4} \eta_{\alpha\beta} F_{\mu\nu}F^{\mu\nu} ) 
       \nonumber \\
     & &  \mbox{} + (\frac{i}{8} \bar{\lambda} \g^\nu 
     [\dsl,\not \! \! A] \chi_\nu +  \mathrm{h.c.})  ] ~,
\label{lagrangian}
\eea
where $\stackrel{\leftrightarrow}{\d} \equiv  \d -
\stackrel{\leftarrow}{\d}$, 
with $\stackrel{\leftarrow}{\d}$ acting on the  function on the left, and
${\cal P}_{R,L} = {1 \over 2} (1 \pm \g_5)$ are the chiral projectors.
The kinetic terms are the canonical ones, with the photon in the Feynman gauge 
and the graviton in the de Donder gauge. 
We use the following notation for particles and their corresponding fields:
\begin{tabbing}
  lepton ~~~~  \= $\rightarrow$ \= $e$ , ~~~~~\= $\Psi$, \hspace{4cm}
  \= sleptons ~~ \= $\rightarrow$ \= $\tilde{e}_{L,R}$ , ~~~ \= $\phi_{L,R}$, \\
  photon   \> $\rightarrow$ \> $\g$ , \> $A_\mu$,  
  \> photino  \> $\rightarrow$ \> $\tilde{\g}$ , \> $\lambda$,  \\
  graviton \> $\rightarrow$ \> $g$ , \> $h_{\mu\nu}$, 
  \> gravitino \> $\rightarrow$ \> $\tilde{g}$ , \> $\chi_\mu$.  
\end{tabbing} 
The lagrangian is written in Minkowski space for easier comparison with 
previous work. However we shall work in Euclidean space as is usual in
DR. The corresponding
Feynman rules are collected in Appendix~A.


\section{Constrained differential renormalization} 
\label{procedure}
The $e^3$ and $e\kappa^2$ corrections to the lepton-lepton-photon vertex 
in SUGRA are 
given by the diagrams in 
Figs.~\ref{figqed} and~\ref{figsugra}, respectively. 
Their expressions contain many
different singular pieces. To obtain renormalized expressions, 
each singular piece must
be substituted by a regular one according to the DR prescription.
It is crucial, however, 
to perform all such renormalizations in a consistent way if the symmetries 
are to be preserved. 
In the literature this is taken care of by imposing certain
relations among the renormalization scales introduced 
in every DR replacement. For
instance, Ward identities in QED require that 
the `logarithmic' and `quadratic'
renormalization scales in the one--loop vacuum polarization 
be equal, while the scales
in the fermion selfenergy, $M_\Sigma$, and the vertex correction, $M_V$, 
must satisfy the equality 
$\log \frac{M_\Sigma}{M_V} = \frac{1}{4}$ ~\cite{FJL}.

Our approach here is to proceed in such a way that  symmetries  are
automatically preserved. 
This can be accomplished if all singular pieces are written in
terms of a minimal set of singular functions, 
which are renormalized afterwards.
This ensures that related singularities are treated in an identical way, 
no matter in which diagram or position they appear. 
Within this scheme, differential renormalization 
should not break the symmetries of the bare 
theory. In Ref.~\cite{SQED} it is shown, for instance,  
that with this approach the vertex Ward 
identity in spinorial QED and scalar QED  is automatically satisfied 
to the one--loop level. Here, the final result will be 
directly compatible with both supersymmetry and $U(1)$ gauge invariance. 

This `constrained' differential renormalization relates diagrams of different 
topology. This is done `separating' the points of diagrams with a 
smaller number of propagators. 
Let us classify  
the diagrams in Figs.~\ref{figqed} and~\ref{figsugra} in two classes according
to their topology:
\begin{itemize}
  \item[{\it i})] 
  {\bf Triangular diagrams}: they are products of three propagators with at most 
  four derivatives. Translation invariance and a
  systematic use of the Leibnitz rule allow to express them in terms of a set
  of functions defined as
  \be
    \T[{\cal O}] \equiv \prop(x) \prop(y) [ {\cal O}^x \prop(x-y) ] ~,
  \label{Tdef}
  \ee
  where $x\equiv x_1-x_3$, $y\equiv x_2-x_3$ and $x_i$ are the coordinates 
  of the triangle vertices. 
  $\prop(x)$ 
  is a scalar propagator 
  and ${\cal O}^x$ is a differential operator acting on $x$
  of \mbox{order~$\leq 4$}. Each of these propagators
  can be either massless or massive. 
  (In the following, with the exception of 
  Eq.~(\ref{Bdef}) below which is analogous to Eq.~(\ref{Tdef}) but for 
  bubble diagrams, 
  we reserve the notation $\prop(x)$ for massless propagators.) 
  In  our calculation  we find T--functions with one massless
  and two massive propagators,
  \be
    \T_1[{\cal O}] \equiv \propm(x) \propm(y) [ {\cal O}^x \prop(x-y) ] ~,
  \label{T1def}
  \ee
  and T--functions with two massless and one massive propagators,
  \be
    \T_2[{\cal O}] \equiv \prop(x) \prop(y) [ {\cal O}^x \propm(x-y) ] ~,
  \label{T2def}
  \ee
  where $\prop(x)=\frac{1}{4\pi^2} \frac{1}{x^2}$ and 
  $\propm(x)=\frac{1}{4\pi^2} \frac{m K_1(mx)}{x}$ fulfill the
  propagator equations
  \bea
    \Box^x \prop(x) & = & - \delta (x)  \label{propeq} ~, \\
    (\Box^x - m^2) \propm(x) & = & - \delta (x)  \label{propmeq} ~,
  \eea
  and $K_i$ are modified Bessel
  functions \cite{Bessel}.
  The mass structure does not affect the singular behaviour. 
  Hence, as far as renormalization is concerned, $\T_1$ and $\T_2$ can be
  considered identical (although some care is needed when both massless and 
  massive DR identities are used, as discussed in Section~\ref{complete}). 
  \item[{\it ii})] 
  {\bf Bubble diagrams}: they are
  products of two propagators and a delta function, and contain at most 
  two derivatives.
  They can be written in terms of a set of functions 
  whose general expression, omitting the index for any massive propagator, is
  \be
    \B_y[{\cal O}] \equiv \delta(y) \prop(x) {\cal O}^x \prop(x)~,
  \label{Bdef}
  \ee
  and analogously for $B_x$, with
  $x \leftrightarrow y$. {${\cal O}^x$} acts on $x$ and 
  is now of order~$\leq 2$. 
  In our calculation B--functions always contain one 
  massive and one massless propagators. Therefore
  the notation
  \bea
    \B_y[{\cal O}] & \equiv & \delta(y) \prop(x) {\cal O}^x \propm(x)~, 
    \label{Gydef} \\
    \B_x[{\cal O}] & \equiv & \delta(x) \prop(y) {\cal O}^y \propm(y)~, 
    \label{Gxdef} 
  \eea
  can be used in what follows.
\end{itemize}

\noindent 
Although the structure of B--functions and T--functions is apparently 
different ---and {\it a priori} they could 
be renormalized differently---, they are in fact related. 
B--functions can be expressed in terms of T--functions 
using the propagator equality~(\ref{propmeq}).
This substitution `separates' points and allows to express a bubble diagram 
as a triangular one. 
In this way 
one is able to express all diagrams (and then all singularities) 
in terms of T--functions only. 
The procedure is 
described in detail in Section~\ref{complete}. There 
it is also proven that all the relevant singular T--functions  can be
written using one single singular function: $\T[\Box]$. Therefore just one
renormalization will be eventually required and only one arbitrary scale 
parameter will have to be introduced. 
Although the full correction to $(g-2)_l$ is finite and does not
depend on the renormalization scale, we calculate the contribution of
each diagram separately in order to compare with
previous results.


\section{A detailed example} \label{example}

In this Section we present in detail the evaluation of the contribution 
to $(g-2)_l$ of diagram D6 in Fig.~\ref{figsugra}. Using the Feynman 
rules in Appendix~A, this vertex correction 
reads (all indices, 
including the $\gamma$--matrix ones, are in Euclidean space)
\bea
  V_\mu^{(6)}(x_1,x_2,x_3) = & & 2 \frac{ie\kappa^2}{4} [ \g_\alpha 
  (\dsl^{x_1} - m) {\cal P}_R \propm (x_1-x_3) 
  \stackrel{\leftrightarrow}{\d}_\mu^{x_3} \propm (x_3-x_2) ] \nonumber \\
  & & \mbox{} \times [ \g_\beta \dsl^{x_2-x_1} \g_\alpha \prop (x_2-x_1)
  {\cal P}_L (\stackrel{\leftarrow}{\dsl}^{x_2} + m) \g_\beta ]~.
\eea
The factor $2$  comes from the fact that the scalar particle propagating
in the diagram can be either $\tilde{e}_L$ or
$\tilde{e}_R$. Translation invariance allows to write $V_\mu^{(6)}$ as a 
function of $x = x_1-x_3$ and $y = x_2-x_3$ only:
\bea
  V_\mu^{(6)}(x_1,x_2,x_3) & = & V_\mu^{(6)}(x,y) \nonumber \\
  & = & \frac{ie\kappa^2}{2} [ \d_\tau^x \prop (x-y) ]  \nonumber \\
  &   & \mbox{} \times [ \g_\alpha (\dsl^x -m) {\cal P}_R 
  \g_\beta \g_\tau \g_\alpha  \nonumber \\
  &   & \mbox{} \times (\dsl^y +m) \g_\beta (\d_\mu^-) (\propm(x) \propm(y)) ],
\eea
where $\d_\rho^- \equiv \d_\rho^y - \d_\rho^x$. The terms containing a 
$\g_5$ do not contribute to $(g-2)_l$ and hereafter will be ignored. 
As explained in Section~\ref{procedure}, all vertex 
corrections can be written in terms of derivatives of T--functions. 
Using the Leibnitz 
rule and the 
properties of the $\gamma$--matrices, we find
\bea 
  \lefteqn{V_\mu^{(6)}(x,y) =} \nonumber \\
  & & i e \kappa^2 
  \{ ( (m^2 - 2 \d^x \cdot \d^y) \d_\mu^- \g_\beta - 2m \d_\mu^- \d_\beta^-)
  \T_1[\d_\beta]  \nonumber   \\
  & & \mbox{}  + 
  ( (2m^2 - 4 \d^x \cdot \d^y) \g_\beta \delta_{\alpha\mu} -4m \d_\beta^- 
  \delta_{\alpha\mu} + 2\d_\mu^- \d_\beta^- \g_\alpha - 
  2m \d_\mu^- \delta_{\alpha\beta} )
  \T_1[\d_\alpha \d_\beta] \nonumber   \\
  & & \mbox{} + (4\d_\beta^- \g_\alpha \delta_{\mu\gamma} - 
  4m \delta_{\alpha\beta} \delta_{\gamma\mu}
  + 2 \d_\mu^- \g_\beta \delta_{\alpha\gamma} ) 
  \T_1[\d_\alpha \d_\beta \d_\gamma] \nonumber  \\
  & & \mbox{} + 4 \g_\beta \T_1[\d_\beta \d_\mu \Box] \} ~. \label{D6deT}
\eea
The external derivatives must be understood  
in the sense of distribution theory according to the DR
prescription: they act `on the left' over test (wave) functions. Of the four 
T--functions in Eq.~(\ref{D6deT}), only $\T_1[\d_\beta]$ is regular 
(~{\it i.e.}, a tempered 
distribution). The other three are singular at $x=y=0$ and must be renormalized
(regularized). This renormalization is done in three steps.

First, the singular T--functions 
are split according to their tensor structure into 
trace and traceless parts: 
\bea
  \T_1[\d_\alpha \d_\beta] & = & \T_1[\d_\alpha \d_\beta - 
  \frac{1}{4} \delta_{\alpha\beta}
     \Box ] + \frac{1}{4} \delta_{\alpha\beta} \T_1[\Box]~,   \label{trace1}  \\
  \T_1[\d_\alpha \d_\beta\ \d_\gamma] & = & 
  \T_1[\d_\alpha \d_\beta\ \d_\gamma\ - \frac{1}{6}
     ( \delta_{\alpha\beta} \d_\gamma + \delta_{\alpha\gamma} \d_\beta +
     \delta_{\beta\gamma} \d_\alpha ) \Box ]   \nonumber \\
     & & \mbox{} + \frac{1}{6} ( \delta_{\alpha\beta} \T_1[\d_\gamma \Box ] +
     \delta_{\alpha\gamma} \T_1[\d_\beta\Box ] + \delta_{\beta\gamma} 
     \T_1[\d_\alpha\Box ])~,   \label{trace2}  \\
   \T_1[\d_\alpha \d_\beta\ \Box ] & = & \T_1[(\d_\alpha \d_\beta - 
   \frac{1}{4} \delta_{\alpha\beta} \Box) \Box ] + \frac{1}{4} 
   \delta_{\alpha\beta}
   \T_1[\Box \Box]~. \label{trace3}
\eea
The traceless parts are two orders less singular. 
Thus, $\T_1[\d_\alpha \d_\beta - \frac{1}{4} \delta_{\alpha\beta} \Box ]$ 
and $\T_1[\d_\alpha \d_\beta\ \d_\gamma- 
\frac{1}{6} ( \delta_{\alpha\beta} \d_\gamma + \delta_{\alpha\gamma} \d_\beta +
\delta_{\beta\gamma} \d_\alpha ) \Box ]$ are finite and 
$\T_1[(\d_\alpha \d_\beta - \frac{1}{4} \delta_{\alpha\beta} \Box) \Box ]$ 
is `logarithmically'
singular. $\T_1[\Box]$ and $\T_1[\Box \Box]$ remain `logarithmically' and
`quadratically' singular, respectively. $\T_1[\d_\alpha \Box]$, which seems
to be `linearly' singular, is in fact `logarithmically' singular 
because it can be written as a function of $\T_1[\Box]$. 
Indeed, using the symmetry of T--functions
under $x \leftrightarrow -y$ interchange (see Appendix~B),
\be
  \T_1[\d_\alpha \Box] = -{1 \over 2} \d_\alpha^- \T_1[\Box]~.
\label{linearT}
\ee

Second, all singular T--functions are expressed as functions of $\T_1[\Box]$ and 
$\T_1[\Box \Box]$ only. Indeed, the other singular T--function left, 
$\T_1[(\d_\alpha \d_\beta - \frac{1}{4} \delta_{\alpha\beta} \Box) \Box ]$, 
can be written
\bea
  \lefteqn{\T_1[(\d_\alpha \d_\beta - \frac{1}{4} \delta_{\alpha\beta}\Box) 
  \Box ] = } \nonumber  \\
  & & [\frac{1}{3} (\d_\alpha^x \d_\beta^x + \d_\alpha^y \d_\beta^y) - 
  \frac{1}{6} (\d_\alpha^x \d_\beta^y + \d_\alpha^y \d_\beta^x) 
  + {1 \over 12} \delta_{\alpha\beta} (\d^x \cdot \d^y - \Box^x-\Box^y)] 
  \T_1[\Box]  \nonumber  \\
  & & + \frac{1}{6} \frac{1}{(4\pi^2)^2} 
  [(\d_\alpha^x + \d_\alpha^y)(\d_\beta^x + \d_\beta^y)
  - \frac{1}{4} \delta_{\alpha \beta} 
  (\d_\rho^x + \d_\rho^y)(\d_\rho^x + \d_\rho^y) ]  \nonumber \\
  & & \mbox{} \times
  [ \delta (x-y) (\frac{m^3 K_0(mx) K_1(mx)}{x} + 
  m^4 (K_0^2(mx) - K_1^2(mx))) ] ~, 
  \label{cuadsintraza}
\eea
as is proven in Appendix~B. Only the first term in the 
{\it r.h.s.} of Eq.~(\ref{cuadsintraza}) is singular.

Third, $\T_1[\Box]$ and $\T_1[\Box \Box]$ are renormalized. 
The singularity of $\T_1[\Box]$
goes as the inverse of the distance to the origin to the fourth power 
and is easily
renormalized. Inserting the propagator equation~(\ref{propeq}), one obtains
\bea
  \T_1[\Box] & = & - [\propm(x) \propm(y)] \delta(x-y) \nonumber \\
          & = & -\frac{1}{(4\pi^2)^2} \frac{m^2 K_1^2(mx)}{x^2} \delta(x-y)~,
\label{k1square}
\eea
and following Ref.~\cite{MassiveDR} the renormalized T-function is
\be 
  \T_1^R[\Box] = -\frac{1}{(4\pi^2)^2} \delta(x-y) [ \frac{1}{2} (\Box-4m^2)
  \frac{m K_0(mx) K_1(mx)}{x} + \pi^2 \log \frac{\bar{M}^2}{m^2} \delta(x) ]~,
\label{renorTbox}
\ee
where $\bar{M} = 2M / \gamma_E$ is an arbitrary scale.
The renormalized $\T_1[\Box\Box]$  can be found in Ref.~\cite{SQED}. 
However it 
does not contribute to $(g-2)_l$ because the corresponding term in 
Eq.~(\ref{D6deT}) is proportional to $\g_\mu$. 
Hence, the renormalization of this correction to $(g-2)_l$
reduces to renormalizing $\T_1[\Box]$ as above. 

Putting everything together we can now evaluate the contribution of D6 to the 
anomalous magnetic moment of the lepton. We substitute Eqs.~(\ref{trace1}
-- \ref{cuadsintraza}) and~(\ref{renorTbox}) into Eq.~(\ref{D6deT}) ignoring
the $\T_1[\Box\Box]$ term. Then we Fourier transform the renormalized vertex 
correction and put the lepton and the photon on their mass--shells. 
The part proportional
to $p_\mu - p'_\mu $, where $p$ and $p'$ are the incoming fermion momenta, 
gives the $(g-2)_l$ contribution of this diagram. 
The Fourier transforms of the regular
terms entering in $(g-2)_l$ are given in Appendix~C. They add to
\be
  (g-2)_l^{D6} = \frac{\kappa^2 m^2}{4 \pi^2} (\frac{4}{3} 
  \log \frac{m^2}{\bar{M}^2}
  + \frac{19}{18}).
\label{D6result}
\ee
In other diagrams the proliferation of terms makes these manipulations rather
cumbersome.
However, a symbolic program developed for this purpose greatly simplifies
the calculation.


\section{The complete calculation}
\label{complete}
In this Section we describe the main aspects of the computation of 
$(g-2)_l$ for all the diagrams in Figs.~\ref{figqed} and~\ref{figsugra}.

{\bf Diagrams DA, DB, D1 and D6}.
The first two diagrams (Fig.~\ref{figqed}) are of order $e^3$, and 
the rest of order $e\kappa ^2$.
In the example above we obtained the contribution of D6 to  $(g-2)_l$ and 
showed how to treat expressions containing $\T_1$--functions. 
In diagram D1 the same set of functions appears, so the procedure 
is completely analogous. In particular, quadratically singular terms 
are again proportional to $\g_\mu$ and do not contribute
to $(g-2)_l$. In fact, the same occurs for all diagrams because 
the quadratically singular functions are scalars containing 
the maximum number of derivatives (four in T's and two in B's), 
so the only Lorentz vectors left are Dirac gammas.
Diagrams DA and DB are even simpler, for no singular terms contribute 
to $(g-2)_l$ once the trace--traceless splitting has been performed.

{\bf Diagrams D4, D5, D9 and D10}.
These diagrams contain two massless and one massive propagators, 
and are written
in terms of $\T_2$--functions. Here three derivatives appear at most, so there
are no quadratically singular terms (not even in the $\g_\mu$ part).
All $\T_2$--functions are reduced to $\T_2[\Box]$ plus regular terms in 
a similar way as we reduced $\T_1$--functions. 
Then the equation~(\ref{propmeq}) is used to write 
this singular function as
\be
  \T_2[\Box] = m^2 \T_2[1] - \frac{1}{(4\pi^2)^2} \frac{1}{x^4} \delta(x-y),
\ee
which together with the DR identity~\cite{FJL} 

\be
  \left. \frac{1}{x^4} \right|^R = -{1 \over 4} \Box \frac{\log x^2 M^2}{x^2}
  \label{typical}
\ee 
allow to define the renormalized function:
\be 
  \T_2^R[\Box] = m^2 \T_2[1] + \frac{1}{4(4\pi^2)^2} \Box \frac{\log x^2
  M^2}{x^2}  \delta(x-y)~.
\label{renorT2box}
\ee
This renormalization is the same as the one in Eq.~(\ref{renorTbox}), so
it involves the same mass scale $M$. Indeed, if 
the function in Eq.~(\ref{k1square}) is expanded in the mass parameter, we obtain
\be
  \frac{m^2 K_1^2(mx)}{x^2} = \frac{1}{x^4} + {\cal R}(m,x) ~, 
\label{massexpansion}
\ee
where ${\cal R}(m,x)$ is of order $m^2$ and regular, and the {\it r.h.s.} of 
Eq.~(\ref{massexpansion}) can be also 
renormalized with the identity~(\ref{typical}). In
Eq.~(\ref{renorTbox}) we used a massive DR identity instead, in order to 
obtain more compact expressions, but we were careful to define the 
renormalized expression in such a way that it agrees with the one 
obtained from massless renormalization of the expansion in 
Eq.~(\ref{massexpansion}). Hence the scales in Eqs.~(\ref{renorTbox}) 
and~(\ref{renorT2box}) are the same.

{\bf Diagrams D2, D3, D6 and D7}.
They can be written in terms of the functions $\B_y$ (D2, D6) and $B_x$ (
D3, D7). As discussed in Section~\ref{procedure}, one must express B--functions 
in terms of T--functions before renormalizing. 
This can be achieved using the equation~(\ref{propmeq})
to `separate' the delta function 
into a propagator, thus obtaining a triangular structure:
\bea
  \B_y[1] & = & \delta(y) \prop(x) \propm(x)  \nonumber \\
     & = & \delta(y) \prop(x-y) \propm(x)  \nonumber \\
     & = & - \Box^y \propm(y) \prop(x-y) \propm(x) + m^2 \propm(y) \prop(x-y) 
     \propm(x) \nonumber \\
     & = & - (\Box^y -m^2) \T_1[1] - 2 \d_\sigma^y \T_1[\d_\sigma]
     - \T_1[\Box]~.
\label{inflation}
\eea
Analogously we obtain:
\bea
  \B_y[\d_\mu] & = & -\d_\mu^x (\Box^y -m^2) \T_1[1] + (\Box^y -m^2) \T_1[\d_\mu]
     - 2\d_\mu^x \d_\sigma^y \T_1[\d_\sigma]  \nonumber  \\
     & & \mbox{} -\d_\mu^x \T_1[\Box] + 2 \d_\sigma^y \T_1[\d_\mu \d_\sigma] +
     \T_1[\d_\mu \Box]~,  \\
  \B_y[\d_\mu \d_\nu] & = & - \d_\mu^x \d_\nu^x (\Box^y -m^2) \T_1[1] +
     \d_\mu^x (\Box^y - m^2) \T_1[\d_\nu] + \d_\nu^x (\Box^y - m^2) 
     \T_1[\d_\mu] \nonumber  \\
     & & \mbox{}-2\d_\mu^x \d_\nu^x \d_\sigma^y \T_1[\d_\sigma] - 
     \d_\mu^x \d_\nu^x \T_1[\Box] - (\Box^y -m^2) \T_1[\d_\mu \d_\nu] \nonumber \\
     & & \mbox{} + 2\d_\mu^x \d_\sigma^y \T_1[\d_\nu \d_\sigma] + 
     2\d_\nu^x \d_\sigma^y \T_1[\d_\mu \d_\sigma] + \d_\mu^x \T_1[\d_\nu \Box] +
     \d_\nu^x \T_1[\d_\mu \Box] \nonumber \\
     & & \mbox{}- 2 \d_\sigma^y \T_1[\d_\mu \d_\nu \d_\sigma] - 
     \T_1[\d_\mu \d_\nu \Box] ~,
\eea
and similar formulae for $B_x$.
Had we na\"{\i}vely renormalized the B-functions independently, we would have 
{\it a priori} no control over the
relation between the renormalization scales of 
triangular diagrams $M_{\T}$ and of bubble diagrams $M_{\B}$, and the choice
$M_{\T}=M_{\B}$ would break SUSY.

Once all the relevant terms in a diagram have been renormalized, one only has 
to extract from the regular expressions the contribution 
to $(g-2)_l$. This is conveniently done by performing a Fourier transform 
and taking the appropriate on--shell limits, as described in 
Appendix~C. 

Let us summarize the steps
we have followed to evaluate $(g-2)_l$:
\begin{enumerate}
  \item All diagrams are written in terms of T-- and B--functions. 
  \item B--functions are rewritten in terms of T--functions.
  \item The singular parts of the T--functions are identified and renormalized.
  \item Finally, the contribution to $(g-2)_l$ is extracted Fourier 
transforming, putting the external leptons on--shell and taking the 
$q^2\rightarrow 0$ limit.
\end{enumerate}
The explicit calculations have been carried out by hand and checked
with a dedicated symbolic program. The necessary equalities are  
collected in the Appendices. 

The resulting contributions of the superQED diagrams are 
\bea
  \mathrm{DA} & \rightarrow & \left( \frac{g-2}{2} \right)^A = 
    \frac{\alpha}{2 \pi} ~,\\
  \mathrm{DB} & \rightarrow & \left( \frac{g-2}{2} \right)^B = 
    - \frac{\alpha}{2 \pi} ~.
\eea 
Note that they are finite, scale independent, and, of course, agree 
with previous results.

The graviton and gravitino corrections to 
$(g-2)_l$ are given in  Table~1, 
together with the dimensional reduction and dimensional regularization 
results.
In next Section we comment on the different contributions 
and compare them. 

At this point,
as the total graviton and gravitino contributions are
finite, one may wonder if it is possible 
to add all the contributions of the different 
diagrams to start with and work with non--singular expressions. 
Actually this requires treating all the diagrams in 
appropriate manner.
Writing all the contributions in terms of T--functions
our procedure allows to obtain a compact result
free of singularities.
We find for the part of the vertex amplitude contributing 
to $(g-2)_l$  
\begin{eqnarray}
      & & \frac{-ie\kappa^2}{2}
      \left\{ \left[ -2m^2 \partial^-_\mu
      \gamma_{\beta} +4 m^3 \delta_{\mu \beta}
      -8 m \partial^-_{\mu} \partial^-_{\beta}  \right] \right.
      \T_1 [ \partial_{\beta} ]    \nonumber \\
      & & \phantom{\frac{-ie\kappa^2}{2}}
      - 8 m \partial^-_{\mu} \T_1 [ \Box ]
      + 8 m^3 \partial^-_{\mu} \T_2 [ 1 ]  \nonumber \\
      & & \phantom{\frac{-ie\kappa^2}{2}}
      \left. + \mbox{} 8m \partial^-_{\mu} \partial^-_{\beta}
      \T_2 [ \partial_{\beta} ] +
      8 m \partial^-_{\mu} \T_2 [ \Box ]
      \right\}.  \label{sumtot}
\end{eqnarray}

\noindent
This expression is regular because the only singular terms are 
those proportional to 
$T_1 \left[ \Box \right]$ and $T_2 \left[ \Box \right]$, which  
have the same singular parts:

\begin{eqnarray}
\T_2 [\Box ] &= & \T_1 [\Box]
- ( \Delta (x)-\Delta_m (x) ) \Delta_m (x) \delta(x-y)  \nonumber \\
& & - ( \Delta (x)-\Delta_m (x) ) \Delta(x) \delta(x-y)
+m^2 \T_2[1],  \label{t2t1}
\end{eqnarray}

\noindent 
and appear with opposite sign. 


\begin{table}
\begin{center}
\begin{tabular}{l l l l}
\hline
  Diagram      & Differential    & Dimensional     &Dimensional \\
               & Renormalization & Reduction       &Regularization \\ \hline \\
                 
D1      & $-\frac{1}{6} \log\left(\frac{\bar{M}^2}{m^2}\right) -\frac{25}{18}$ 
        & $\frac{1}{3} \frac{1}{n-4}-\frac{29}{18}$
        & $\frac{1}{3}\frac{1}{n-4}-\frac{61}{36}$ \\  \\
D2+D3   & $-\frac{11}{6} \log \left(\frac{\bar{M}^2}{m^2}\right)-\frac{11}{18}$   
        & $\frac{11}{3} \frac{1}{n-4}-\frac{35}{9}$
        & $\frac{11}{3} \frac{1}{n-4}-\frac{32}{9}$ \\ \\
D4+D5   & $ 2  \log \left(\frac{\bar{M}^2}{m^2}\right) +1$     
        &$-4\frac{1}{n-4}+6$
        &$-4\frac{1}{n-4}+7$ \\	 \\
Graviton & & &  \\     
(D1+D2+D3+D4+D5) &$-1$              &1/2          &7/4 \\	 \\	\hline\\
D6         &$-\frac{4}{3} \log \left(\frac{\bar{M}^2}{m^2}\right) +\frac{19}{18}$ 
           &$\frac{8}{3}  \frac{1}{n-4}-\frac{37}{18}$
           &$\frac{8}{3}  \frac{1}{n-4}-\frac{55}{18}$ \\   \\
D7+D8      &$-\frac{2}{3} \log \left(\frac{\bar{M}^2}{m^2}\right)+\frac{17}{18}$ 
           &$\frac{4}{3}  \frac{1}{n-4}-\frac{4}{9}$
           &$\frac{4}{3}  \frac{1}{n-4}-\frac{13}{9}$ \\	\\
D9+D10     &$ 2  \log \left(\frac{\bar{M}^2}{m^2}\right) -1$     
           &$-4\frac{1}{n-4}+2$
           &$-4\frac{1}{n-4}+4$ \\ \\
Gravitino  &  &  &   \\  
(D6+D7+D8+D9+D10)  &1             &$-1/2$          &$-1/2$ \\ \\ \hline \\
TOTAL     &   &   & \\
(Graviton+Gravitino) & 0 &0 &5/4 \\ \\ \hline
\end{tabular}
\end{center}
\caption{Contributions of the diagrams in Fig.~2 
to $\left( \frac{g-2}{2} \right)_l$ in units of 
$\frac{G_N m^2}{\pi}$, obtained with DR, dimensional reduction
and  dimensional regularization.}
\end{table}


Another important remark, further discussed in Appendix~C,
is the infrared divergent behaviour of the Fourier transforms of 
the $\T_2$--functions in the static limit $q^2 \rightarrow 0$. 
It can be shown, however, 
that these divergencies cancel out in the  
sum Eq.~(\ref{sumtot}). In fact we have checked that all 
diagrams, with the external leptons on--shell, are well--behaved 
in this limit. 

We have also verified the absence of gauge non--invariant
terms, proportional to the photon momentum $q_\mu$,
when the leptons are on their mass shell. This corresponds
in coordinate space to a symmetry under the $x \leftrightarrow -y$
interchange, 
which is preserved throughout the calculation.



\section{Conclusions} \label{conclusions}
Our explicit results are summarized in  Table~1. 
The contributions
to $(g-2)_l$ of the graviton and gravitino sectors 
in differential renormalization are 
separately finite and independent of the scale $\bar{M}$. 
In fact, in our procedure the logarithms of $\bar{M}$ 
keep track of the singularities  and their sum  
is shown to vanish. Notice  that these logarithms are proportional
to the infinities of the regularization  
methods in momentum space, as  expected at one loop. 
Obviously, finite parts
may be different and are responsible for the preservation or the breaking
of the symmetries in the problem. In differential renormalization
the total contribution
cancels out as dictated by supersymmetry. In the dimensional 
reduction scheme the total contribution also vanishes.
This is not the case in 
dimensional regularization as this scheme is known to break 
supersymmetry.

Note that the values of the $(g-2)_l$ contribution of the graviton
and gravitino sectors are different in differential renormalization $(1,-1)$ 
and in dimensional reduction $(-1/2, 1/2)$, although
both schemes preserve supersymmetry.
These  contributions  appear to be regularization
dependent. This
ambiguity is related to the presence of linear singularities (divergences) 
at one loop in SUGRA.
All that can be said is that global supersymmetry ensures
a vanishing value of $(g-2)_l$ and a cancellation of the ambiguities
between the  graviton and gravitino sectors.
In Ref. \cite{Wilcox} 
yet another finite value for the graviton contribution to $(g-2)_l$, 
$-13/12 (G m^2/\pi)$, was obtained using source theory
techniques. The situation is that in four different schemes
the graviton contribution is different. 
The Pauli term is  allowed  in the gravity lagrangian 
and its one--loop corrections are scheme dependent in such a 
non--renormalizable theory. 

An important point developed in our work is the exact
maintenance of 
symmetries within the differential renormalization scheme.
Direct renormalization with different scales requires 
enforcing the relevant symmetries (gauge and supersymmetry) with 
the corresponding Ward identities at the end of
the calculation. In our approach all expressions have been 
handled in a symmetric way.
We have reduced all singularities to a minimal set of independent 
functions. This was done relating diagrams of 
different topology with the technique of point separation. 
This procedure ensures that singularities are consistently
treated. Then we obtain a result which is supersymmetric 
and compatible with U(1) gauge invariance. 

\subsection*{Aknowledgements}
F.A. thanks R. Stora for discussions. 
The symbolic program MATHEMATICA has been extensively used.
The figures have been produced with the Feyndiag package.
This work has been supported by CICYT contract number AEN96--1672,
Junta de Andalucia 1201,
and European Network Contracts CHRX-CT-92-0004 and ERBCHBICT941777 (R.M.T.).


\appendix
\newpage

\section*{Appendix A}
\begin{figure}[h]
\setlength{\unitlength}{0.68cm}
\begin{center}
\begin{picture}(10,19)
\epsfxsize=6.8cm
\put(0,0){\epsfig{file=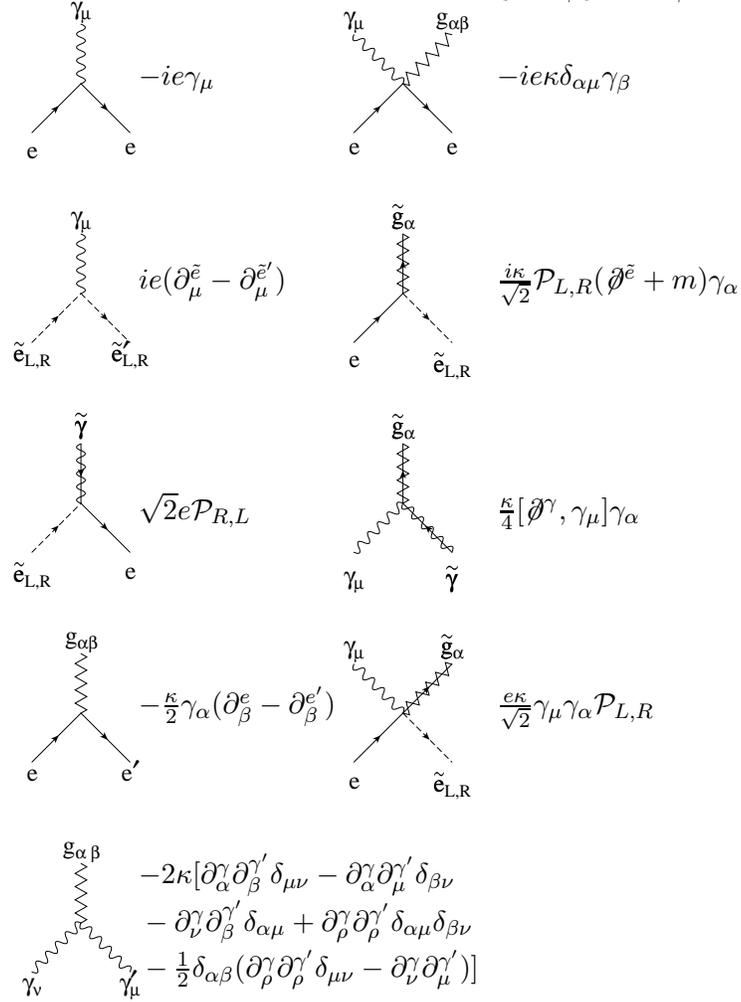}}
\put(3,18){$-i e \gamma_\mu$}
\put(3,14){$i e (\partial_\mu^{\tilde{e}} -
           \partial_\mu^{\tilde{e}^\prime})$}
\put(3,9.5){$\sqrt{2} e {\cal P}_{R,L}$}
\put(3,5.7){$- \frac{\kappa}{2} \gamma_\alpha 
           (\partial_\beta^e -
           \partial_\beta^{e^\prime})$}
\put(10,18){$-ie\kappa \delta_{\alpha\mu} \gamma_\beta $}
\put(3,1.5){\parbox{5cm}{$-2 \kappa [ \partial_\alpha^\gamma 
          \partial_\beta^{\gamma^\prime} 
          \delta_{\mu\nu} - \partial_\alpha^\gamma \partial_\mu^{\gamma^\prime}
          \delta_{\beta\nu}$ \\
          $\mbox{} - \partial_\nu^\gamma \partial_\beta^{\gamma^\prime} 
          \delta_{\alpha\mu}
          + \partial_\rho^\gamma \partial_\rho^{\gamma^\prime} \delta_{\alpha\mu}
          \delta_{\beta\nu}$ \\
          $\mbox{} - \frac{1}{2} \delta_{\alpha\beta} 
          (\partial_\rho^\gamma \partial_\rho^{\gamma^\prime} \delta_{\mu\nu} 
          - \partial_\nu^\gamma \partial_\mu^{\gamma^\prime})]$}}
\put(10,14){$\frac{i \kappa}{\sqrt{2}} {\cal P}_{L,R} (\not \! 
          \partial^{\tilde{e}} + m)  \gamma_\alpha$}
\put(10,9.5){$\frac{\kappa}{4} [\not \! \partial^\gamma,\gamma_\mu] \gamma_\alpha$}
\put(10,5.7){$\frac{e\kappa}{\sqrt{2}} \gamma_\mu \gamma_\alpha {\cal P}_{L,R}$}
\end{picture}
\end{center}
\caption{Feynman rules for vertices. ${\cal P}_{R,L}=\frac{1}{2}(1 \pm \gamma_5)$ 
         are the chiral projectors. All derivatives are  
         with respect to the vertex space-time point. The superscripts 
         indicate the field they are
         acting on.  The rules 
         for diagrams with opposite charge and fermion number arrows are obtained 
         from these by the transformation 
         $\mbox{FR} \rightarrow \gamma_5 \mbox{FR}^\dagger 
         \gamma_5$~[25].
         \label{figFeynrulesver}}
\end{figure}

\begin{figure}[ht]
\setlength{\unitlength}{1cm}
\begin{center}
\begin{picture}(6,12)
\put(0,0){\epsfig{file=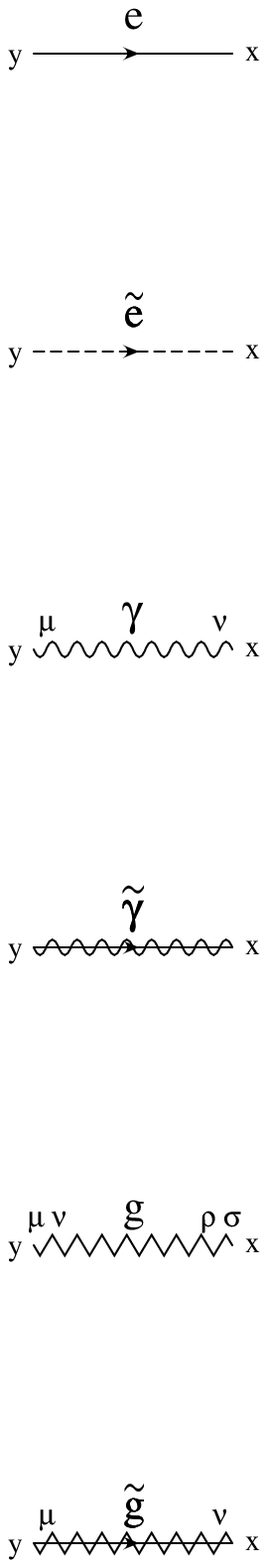}}

\put(3.5,15.6){$- (\not \! \partial^x -m) \Delta_m(x-y)$}
\put(3.5,12.45){$\Delta_m(x-y)$}
\put(3.5,9.6){$\delta_{\mu\nu} \Delta(x-y)$}
\put(3.5,6.5){$- \not \! \partial^x \Delta(x-y)$}
\put(3.5,3.5){\parbox{5cm}{$\frac{1}{2} (\delta_{\mu\rho}\delta_{\nu\sigma}+
             \delta_{\mu\sigma}\delta_{\nu\rho}$ \\
             $\mbox{} - \delta_{\mu\nu}\delta_{\rho\sigma}) \Delta(x-y) $}}
\put(3.5,0.45){$\frac{1}{2} \gamma_\nu \not \! \partial^x \gamma_{\mu} 
              \Delta(x-y)$}
\end{picture}
\end{center}
\caption{Feynman rules for propagators. $\Delta(x)=\frac{1}{4\pi^2} 
         \frac{1}{x^2}$ and 
         $\Delta_m(x)=\frac{1}{4\pi^2} \frac{m K_1(mx)}{x}$ are the massless 
         and massive scalar propagators. Here the superscript on the 
         derivatives refers to the space-time point. 
         \label{figFeynrulesprop}}
\end{figure}

\setcounter{section}{1} 
\label{feynmanrules}
In this Appendix we give the Feynman rules 
(Figs.~\ref{figFeynrulesver} and~\ref{figFeynrulesprop}) for 
the lagrangian~(\ref{lag1} -- \ref{lagrangian}). 
They are written in Euclidean
space with the convention $\{\g_\alpha,\g_\beta\} = 2 \delta_{\alpha\beta}$.

\newpage



\section*{Appendix B}
\setcounter{section}{2}
\label{Appreduction}

In this Appendix we show how $\T_1 [\d_\alpha \Box]$ and 
$\T_1 [(\d_\alpha \d_\beta - {1 \over 4} \delta_{\alpha\beta} \Box)
\Box ]$ can be expressed in terms of the logarithmically singular 
function $\T_1 [\Box]$.
Let us first deal with the linearly singular $\T_1 [\d_\alpha \Box]$: 
\bea
  \T_1 [\d_\alpha \Box] & = & \propm(x) \propm(y) \d_\alpha^x \Box^x 
     \prop(x-y)  \nonumber  \\
  & = & - \propm(x) \propm(y) \d_\alpha^x \delta(x-y)  \nonumber  \\
  & = & - \d_\alpha^x [ \propm(x) \propm(y) \delta(x-y) ] +
     [ (\d_\alpha^x  \propm(x)) \propm(y) ] \delta(x-y)  \nonumber  \\
  & = & \d_\alpha^x \T_1 [\Box] + [ (\d_\alpha^x \propm(x)) \propm(x) ] 
     \delta(x-y) ~.
\label{linx}
\eea
On the other hand, 
\bea
  \T_1 [\d_\alpha \Box] & = & - \propm(x) \propm(y) \d_\alpha^y \Box^x 
     \prop(x-y)  \nonumber  \\
  & = & - \d_\alpha^y \T_1 [\Box] - [\propm(x) (\d_\alpha^x \propm(x)) ] 
     \delta(x-y) ~.
\label{liny}
\eea
So, combining both expressions 
\be
  \T_1 [\d_\alpha \Box] = {1 \over 2}(\d_\alpha^x - \d_\alpha^y) \T_1 
  [\Box] ~. 
\label{linsum}
\ee
With the `integration by parts' prescription the linear singularity has 
been reduced to a logarithmic one.
For $\T_1 [(\d_\alpha \d_\beta - {1 \over 4} \delta_{\alpha\beta} \Box)
\Box ]$ we have 
\bea
  \T_1 [(\d_\alpha \d_\beta - {1 \over 4} \delta_{\alpha\beta} \Box)
  \Box ] & = & 
  - {1 \over 2} (\d_\alpha^x \d_\beta^x + \d_\alpha^y \d_\beta^y -
  {1 \over 4} \delta_{\alpha\beta} (\Box^x + \Box^y) ) 
  [( \propm(x))^2 \delta(x-y)]  \nonumber \\
  & & \mbox{} + [\d_\alpha^x \propm(x) \d_\beta^x \propm(x) - {1 \over 4}
  \delta_{\alpha\beta} \d_\rho^x \propm(x) \d_\rho^x \propm(x) ] \nonumber \\ 
  & & \mbox{} \times \delta(x-y) ~,
\eea
where the previous result has been used. Using recurrence relations among 
modified Bessel functions, the second term in the {\it r.h.s.} can be written
\bea
  \lefteqn{\frac{1}{(4 \pi^2)^2} (\frac{x_\alpha x_\beta}{x^2} - {1 \over 4}
  \delta_{\alpha\beta}) \frac{m^4 K_2^2 (mx)}{x^2} \delta(x-y)} \nonumber \\
  & = & \frac{1}{(4 \pi^2)^2} {1 \over 6} \delta(x-y) [ (\d_\alpha^x \d_\beta^x 
  -{1 \over 4} \delta_{\alpha\beta} \Box) (\frac{m^2 K_1^2 (mx)}{x^2}
  + \frac{m^3 K_0 (mx) K_1 (mx)}{x} \nonumber  \\ 
  & & \mbox{} + m^4 (K_0^2 (mx) - K_1^2 (mx) ) ] \nonumber  \\
  & = & \frac{1}{(4 \pi^2)^2} {1 \over 6} [ (\d_\alpha^x + \d_\alpha^y)
  (\d_\beta^x + \d_\beta^y) - {1 \over 4} \delta_{\alpha\beta}
  (\d_\rho^x + \d_\rho^y) (\d_\rho^x + \d_\rho^y)]  \nonumber \\
  & & \mbox{} \times [\delta (x-y) (\frac{m^2 K_1^2 (mx)}{x^2} +
  \frac{m^3 K_0 (mx) K_1 (mx)}{x} \nonumber \\
  & & \mbox{}+ m^4 (K_0^2 (mx) - K_1^2 (mx)))],  ~
\label{formula}
\eea
where the identity 
\be
[\d_\mu^x f(x)] \delta(x-y) = (\d_\mu^x + \d_\mu^y) [f(x) \delta(x-y)]
\ee
has been used to obtain the last equality.
Only the term with $\frac{m^2 K_1^2 (mx)}{x^2}$ 
in Eq.~(\ref{formula})  
is singular and equal to 
\be
  - {1 \over 6} [ (\d_\alpha^x + \d_\alpha^y)
  (\d_\beta^x + \d_\beta^y) - {1 \over 4} \delta_{\alpha\beta}
  (\d_\rho^x + \d_\rho^y) (\d_\rho^x + \d_\rho^y)] \T_1 [\Box]  ~.
\ee
Thus we finally obtain
\bea
  \lefteqn{\T_1[(\d_\alpha \d_\beta - \frac{1}{4} \delta_{\alpha\beta}\Box) 
  \Box ] = } \nonumber  \\
  & & [\frac{1}{3} (\d_\alpha^x \d_\beta^x + \d_\alpha^y \d_\beta^y) -\frac{1}{6}
  (\d_\alpha^x \d_\beta^y + \d_\alpha^y \d_\beta^x) 
  + {1 \over 12} \delta_{\alpha\beta} (\d^x \cdot \d^y - \Box^x-\Box^y)] 
  \T_1[\Box]  \nonumber  \\
  & & + \frac{1}{6} \frac{1}{(4\pi^2)^2} [(\d_\alpha^x + 
  \d_\alpha^y)(\d_\beta^x + \d_\beta^y)
  - \frac{1}{4} \delta_{\alpha \beta} 
  (\d_\rho^x + \d_\rho^y)(\d_\rho^x + \d_\rho^y) ]  \nonumber \\
  & & \mbox{} \times
  [ \delta (x-y) (\frac{m^3 K_0(mx) K_1(mx)}{x} + m^4 (K_0^2(mx) - K_1^2(mx))) ] ~.
  \label{appquadtraceless}
\eea


\section*{Appendix C}
\setcounter{section}{3}
\label{AppFourier}

In order to extract the contribution to the anomalous magnetic moment 
of a given graph from its full renormalized expression, external fields 
must be on--shell. External derivatives in
1PI graphs act directly by parts on the external fields, 
so the Dirac or Maxwell equations can be
used straightforwardly. When internal derivatives are present, however, 
the situation is more
involved due to the noncommutative character of the derivatives. 
A simple procedure to follow is to
perform a Fourier transform of the whole graph and deal with momenta, which 
are commuting objects.
Notice that all Fourier transforms are ultraviolet 
convergent since the singular pieces have already been
renormalized. On--shell conditions can then be readily imposed and the 
$(g-2)_l$ contribution identified.

The Fourier transform of a distribution $f(x,y)$ is
\be
  \hat{f}(p,p^\prime)=\int d^4 x\, d^4 y \, e^{ip \cdot x} 
  e^{i p^\prime \cdot y} f(x,y) ~,
\label{fourierdef}
\ee
where $p$ and $p^\prime$ are the incoming momenta of the external leptons 
and $q=p+p^\prime$ is the outgoing momentum of the external photon.
External derivatives yield
\be 
  \d_\mu^x \rightarrow -i p_\mu ~;~\d_\mu^y \rightarrow -i p_\mu^\prime 
\ee
and we are left with the Fourier transforms of T--functions. When 
computing them we distinguish the `originally regular' ones from the 
`renormalized' ones (which, of course, are also regular, 
but involve different kinds of functions). The originally regular T--functions are
products of three propagators with some inner derivatives. Their Fourier 
transform is a convolution of momentum space propagators with the corresponding 
internal momenta. For $\T_1$--functions:
\be
  \hat{\T}_1[{\cal O}(\d)] = \I[{\cal O}(\d \rightarrow ik)] ~,
\ee
where 
\bea
  & & \I[f(k)] \equiv \int \frac{d^4 k}{(2\pi)^4}\, \hat{\Delta}_m(p+k) 
      \hat{\Delta}_m(p^\prime-k) \hat{\prop}(k)~, \\
  & & \hat{\prop}(k)=\frac{1}{k^2} ~,~~\hat{\Delta}_m(k)= \frac{1}{k^2+m^2}~.
\eea
Thus, {\it e.g.}\/,
\be
  \hat{\T}_1 [\d_\alpha \d_\beta - {1 \over 4} \delta_{\alpha\beta} \Box] = 
  - \I[k_\alpha k_\beta - {1 \over 4} k^2 \delta_{\alpha\beta}].
\ee
The integrals $\I[f(k)]$ appear in one--loop 
momentum space calculations and can be evaluated with
standard techniques such as Passarino-Veltman reduction and Feynman 
parametrization.
In our case, since we are interested in the static limit
\be
  p^2=p^{\prime 2}=-m^2~,~~q^2 \rightarrow 0~,
\label{staticlimit}
\ee
it is convenient to impose these conditions before computing the integrals, 
which then become much simpler \cite{onshellints}.

Formally, the Fourier transform of originally regular $\T_2$-functions is 
identical:
\be
  \hat{\T}_2[{\cal O}(\d)] = \J[{\cal O}(\d \rightarrow ik)] ~,
\ee
where
\be
  \J[f(k)] \equiv \int \frac{d^4 k}{(2\pi)^4}\, \hat{\prop}(p+k) 
      \hat{\prop}(p^\prime-k) \hat{\Delta}_m(k)~.
\ee
However the presence of two massless propagators makes the static limit of these 
integrals
infrared divergent\footnote[1]{$\hat{\T}_1[1]$ is also 
infrared divergent when $q^2 \rightarrow 0$, but
it does not appear in the $(g-2)_l$ parts.}, so one must keep $q^2 \neq 0$. 
It is gratifying to note that all the infrared 
divergencies cancel out in the $(g-2)_l$ part of each diagram, rendering the 
limit $q^2 \rightarrow 0$ finite.

Renormalized expressions of T--functions always contain a delta function, 
$\delta(x-y)$.
Their Fourier transforms become integrals over one (four-dimensional) variable:
\be
  \int d^4 x \, d^4 y \, e^{ip\cdot x} e^{i p^\prime \cdot y} [g(x) 
  \delta(x-y)] =
  \int d^4 x \,e^{i q\cdot x} g(x)~,
\ee
with $q_\mu=p_\mu + p_\mu^\prime$. Hence we only need the Fourier transforms 
of certain functions of one  variable containing logarithms or modified 
Bessel functions. Derivatives, if present, 
are trivially pulled out yielding momenta $q$. The transform of 
$\log M^2 x^2 \, /\, x^2$ was
given in \cite{FJL}:
\be
  \int d^4 x \, e^{iq\cdot x} \frac{\log M^2 x^2}{x^2} = 
  -\frac{4\pi^2}{q^2} \log \frac{q^2}{\bar{M}^2} ~.
\label{masslessfourier}
\ee
Note that Eq.~(\ref{masslessfourier}) is divergent when $q^2 \rightarrow 0$. 
In the computation of $(g-2)_l$, 
infrared divergencies coming from originally regular pieces and 
from renormalized pieces cancel
separately.
To obtain the Fourier transforms of expressions with modified Bessel functions, 
the following recurrence relations in four dimensions prove useful:
\bea
  \Box \frac{K_1(x)}{x} & = & \frac{K_1(x)}{x}- 4 \pi^2 \delta(x), \\
  \Box K_2(x) & = & K_0(x) - 8 \pi^2 \delta(x), \\
  \Box K_0(x) & = & K_0(x)- 2 \frac{K_1(x)}{x}, \\
  \Box K_1^2(x) & = & 2 (K_0^2(x) + K_1^2(x)) - 4 \pi^2 \delta(x), \\
  \Box (K_1^2(x)-K_0^2(x)) & = & 4 \frac{K_0(x) K_1(x)}{x} - 4\pi^2 \delta(x)~.
\eea
All these expressions are tempered distributions.
The transforms we need are
\bea
  \int d^4 x \, e^{iq\cdot x} \frac{mK_1(mx)}{x} & = & 4 \pi^2 
    \hat{\Delta}_m(q), \\
  \int d^4 x \, e^{iq\cdot x} K_0(mx) & = & 8\pi^2 [\hat{\Delta}_m(q)]^2, \\
  \int d^4 x \, e^{iq\cdot x} \frac{m K_0(mx) K_1(mx)}{x} & = &
    32 \pi^4 \int \frac{d^4 k}{(2\pi)^4} \, \hat{\Delta}_m(q-k) 
    [\hat{\Delta}_m(q)]^2 \nonumber \\
    & \stackrel{q^2 \rightarrow 0}{=} & \frac{\pi^2}{m^2}, \\
  \int d^4 x \, e^{iq\cdot x} m^2(K_1^2(mx) - K_0^2(mx)) & = &
    -128 \pi^4 \frac{m^2}{q^2} 
    \int \frac{d^4 k}{(2\pi)^4} \, \hat{\Delta}_m(q-k) [\hat{\Delta}_m(q)]^2 
    \nonumber \\
    & & \mbox{} + \frac{4\pi^2}{q^2} \nonumber \\
    & \stackrel{q^2 \rightarrow 0}{=} & {2 \over 3} \frac{\pi^2}{m^2}.
\eea

\noindent Finally, the momentum space expressions allow for direct use of 
the Dirac equation 
(\mbox{$\not \! p \rightarrow -im, ~\not \!p^\prime \rightarrow im$}), and the 
$(g-2)_l$  parts are easily recognized, 
for they are proportional to $p_\mu - p_\mu^\prime$.


\end{document}